\documentstyle[12pt]{article}

\newcommand{\be}{\begin{eqnarray}}
\newcommand{\en}{\end{eqnarray}}
\newcommand{\ka}{\kappa}
\newcommand{\ga}{\gamma}
\newcommand{\ba}{\beta}
\newcommand{\al}{\alpha}
\newcommand{\gb}{\bar{g}}
\newcommand{\ab}{\bar{\al}}
\newcommand{\ben}{\begin{eqnarray*}}
\newcommand{\enn}{\end{eqnarray*}}

\begin{document}
\begin{titlepage}	
\begin{flushright}
EFI 96-09 \\
hep-th/9604021 

\end{flushright}

\begin{center}
\vskip 0.3truein

{\large{\bf{ASYMPTOTIC LIMITS AND SUM RULES}}}
\vskip 0.15truein

{\large{\bf{FOR THE QUARK PROPAGATOR}}}
\footnote{Work supported in part by the National
Science Foundation, Grant PHY 91-23780. \\
E-mail: oehme@control.uchicago.edu}
\vskip 0.6truein

{\large{ Reinhard Oehme and Wentao Xu}}
\vskip 0.2truein

Enrico Fermi Institute and Department of Physics

University of Chicago

Chicago, Illinois 60637, USA
\end{center}
\vskip 0.5truein

\centerline{Abstract}

\bigskip

For the structure functions of the quark propagator, the asymptotic
behavior is obtained for general, linear, covariant gauges, and in
all directions of the complex $k^2$-plane. Asymptotic freedom is assumed.
Corresponding previous results for the gauge field propagator are
important in the derivation. Except for coefficients, the leading 
asymptotic terms are determined by one-loop or by two-loop information, 
and are gauge independent. Various sum rules are derived. 

\end{titlepage}
\newpage

\bigskip
\setcounter{equation}{0}

In a previous paper \cite{ox}, we have obtained the asymptotic behavior
of the gauge field propagator in general, covariant, linear gauges, 
and for all directions in the complex $k^2$-plane. Given asymptotic   
freedom of the theory, we found interesting sum rules \cite{ox,oz}. Except for 
coefficients, the functional form of the leading asymptotic terms is 
gauge independent, and determined by the one-loop anomalous dimension 
coefficient $\gamma_{00}$ and the corresponding $\beta$-function coefficient 
$\beta_{0}$. The asymptotic properties, and the related superconvergence 
relations, are of interest for the problem of confinement \cite{o,np}.
If applied to certain models of $N=1$ supersymmetric gauge theories \cite{of,on}, 
the results obtained concerning the confinement of gauge field quanta,  
on the basis of superconvergence, are in agreement 
with those derived from duality \cite{ws,s}. In particular, there are interesting 
relations between the one-loop anomalous dimension coefficient $\gamma_{00}$
of the SUSY gauge theory and the one-loop $\beta$-function coefficient of
the dual map \cite{on}. These relations underline the relevance of the sign
of the coefficient $\gamma_{00}$ for supercovergence, duality and confinement.

It is the purpose of this note to present general results for the quark 
propagator. We obtain 
the asymptotic terms of the structure functions in general, linear, covariant 
gauges. Except for the case of the Landau gauge, 
which has been considered before \cite{gw,oz}, the previous results for the 
gauge field propagator play an important r\^{o}le in the derivation. Again only 
anomalous dimension and $\beta$-function coefficients in lowest non-vanishing 
order of perturbation theory are relevant, and gauge dependence is limited.

Our results are based upon general principles only. We use Lorentz-covariance, 
and minimal spectral condition as formulated in the state space of indefinite metric 
of the gauge theories considered \cite{ko}. For some results, 
we assume that the exact amplitudes 
connect with the weak coupling perturbative expressions in the limit of vanishing 
coupling, at least as far as the leading term is concerned. 
Our main tool is the renormalization group. We do not consider 
topological aspects of the theory. They should not be of importance within our
general framework. 

\medskip

In this report, we concentrate on the derivation of the asymptotic 
terms. Possible applications, in particular generalizations to SUSY 
theories and their dual maps, will be explored elsewhere. One may hope
to find further correlations between the phase structure of the SUSY
systems, as inferred from duality, and specific properties of the propagators.
Since we are interested in asymptotic expressions, it is sufficient to consider 
theories with no external mass parameters in the action. Possible non-perturbative 
mass generation is not excluded.

We write the ``quark'' propagator 
\be
-iS_{F}^{\al\ba}(k) = \int d^4x \exp^{ik\cdot x}
\langle 0|T\psi^\al(x)\bar{\psi}^\ba(0)|0\rangle
\label{1}
\en
in the form
\be
-iS_F(k) = A(k^2)\sqrt{-k^2} + B(k^2)\ga\cdot k ~,
\label{2}
\en
where gauge and spinor indices have been suppressed. 
With $\ka^2 < 0$ as the normalization point, we have the renormalization group equation
\be
\psi(x, g',\al',\ka'^2) = \sqrt{Z_2}\psi(x,g,\al,\ka^2)~,
\label{3}
\en
and corresponding relations for other fields.
Here 
\be
Z_2 = Z_2\left(\frac{\ka'^2}{\ka^2},g,\al\right), ~~~~ 
g' = \gb\left(\frac{\ka'^2}{\ka^2},g\right) ,\nonumber\\
\al'=\bar{\al}\left(\frac{\ka'^2}{\ka^2},g,\al\right)=\al R^{-1}
\left(\frac{\ka'^2}{\ka^2},g,\al\right) ,
\label{4}
\en
where $R = -k^2D(k^2, \ka^2, g,\al)$ being the structure function of the 
transverse gauge field propagator with the normalization $R(1,g,\al) = 1$.
It is convenient to introduce dimensionless structure function by 
\be
-k^2A(k^2,\ka^2,g,\al) = S\left(\frac{k^2}{\ka^2},g,\al\right) ,~~
-k^2B(k^2,\ka^2,g,2) = T\left(\frac{k^2}{\ka^2},g,\al\right) .
\label{5}
\en
Here we chose the normalization
\be
T(1,g,\al) = 1 .
\label{6}
\en
With Eqs. (\ref{1}), (\ref{3}) and (\ref{6}), we can write the renormalization
group equations for the functions $T$ and $S$ in the form
\be
T\left(\frac{k^2}{\ka^2},g,\al\right) &=& T\left(\frac{\ka'^2}
{\ka^2},g,\al\right)T\left(\frac{k^2}{\ka'^2},g',\al'\right) ,
\label{7}
\en
\be
S\left(\frac{k^2}{\ka^2},g,\al\right) &=& T\left(\frac{\ka'^2}
{\ka^2},g,\al\right)S\left(\frac{k^2}{\ka'^2},g',\al'\right) ,
\label{8}  
\en
because our normalization implies $Z_2^{-1} = T$ and $Z_3^{-1} = R$. The  
primed parameters $g'$ and $\al'$ are given in Eq.(\ref{4}).
In a situation, where the theory has unbroken chiral symmetry, the structure 
function $A$ vanishes identically. But if we allow for the possible non-perturbative 
mass generation, we expect a nonzero function, which vanishes in the perturbative limit.

As a consequence of Lorentz covariance and simple spectral conditions 
formulated in the state space of indefinite metric, it follows that the 
structure functions (distributions) $B(k^2+i0)$ and $A(k^2+i0)$ are boundary values of 
corresponding analytic functions, which are regular in the cut $k^2$-plane 
with branch lines along the positive real $k^2$-axis. In the following, 
we first use the renormalization group in order to obtain the asymptotic 
behavior of these functions for $k^2\rightarrow -\infty$ along the negative 
real $k^2-$axis. We then generalize the results to all directions in the 
complex $k^2$-plane \cite{ox,oz}. For this purpose, we consider Eqs.(\ref{7},\ref{8}) with 
$\ka'^2=-|k^2|$, $k^2=-|k^2|e^{i\phi}, |\phi| \leq \pi$, and find 
\be
T\left(\frac{k^2}{\ka^2},g,\al\right) = T\left(\left|\frac{k^2}{\ka^2}\right|,g,\al\right)
T(e^{i\phi},\bar g,\bar{\al}), 
\label{9}
\en
where $\gb = \gb(|\frac{k^2}{\ka^2}|,g)$, and $\bar\al = \bar\al(|\frac{k^2}{\ka^2}|,g,\al)$.
There is a corresponding expression for the function S, but in the following, we
first discuss only T.

Assuming that the exact structure functions approach their perturbative 
limits for $g^2\rightarrow0$, at least as far as the leading terms are concerned,
we have for vanishing coupling:
\begin{eqnarray}
T\left(\frac{k^2}{\ka^2},g,\al\right) &\simeq& 1+g^2\ga_{F01}
\al\ln\frac{k^2}{\ka^2}+\cdots\nonumber\\
& &\hspace{2in}\mbox{for}~~ \al \neq 0\nonumber\\
T\left(\frac{k^2}{\ka^2},g,\al\right) &\simeq& 1+g^4\ga_{F01}
\ln\frac{k^2}{\ka^2}+\cdots\nonumber\\
& &\hspace{2in}\mbox{for}~~ \al = 0
\label{10}
\end{eqnarray}
In these equations, the anomalous dimension coefficients of the quark field are defined by
$\ga_F(g^2,\al)=\ga_{F0}(\al)g^2+\ga_{F1}(\al)g^4 + \cdots$ , with 
$\ga_{F0}(\al)=\ga_{F00}+\al\ga_{F01}$, $\ga_{F1}(\al)=\ga_{F10}+\al\ga_{F11}+
\al^2\ga_{F12}$. We consider here QCD, or similar theories, so that we always
have $\ga_{F00}\equiv 0$ and $(16\pi^2)\ga_{F01}=C_2(R)>0$, with $C_2(R)=4/3$
in QCD. For SUSY theories, on the other hand, we generally
find $\ga_{F00}\neq 0$ for the Fermi field in the Wess-Zumino representation.
These theories will be discussed elsewhere.  
Since $\gb^2(u,g)\simeq(-\ba_0\ln u)^{-1}$ and $\bar{\al}(u,g,\al)=\al R^{-1}(u,g,\al)
\simeq 0$ or $\al_0 = -\ga_{00}/\ga_{01}$ for $u\rightarrow\infty$, $\al \geq 0 $,
and in view of the analytic properties of the structure functions, we see that 
Eq.(\ref{9}) gives the asymptotic behavior in all directions $\phi$ in terms 
of the limit for real $k^2\rightarrow-\infty$.

In order to make use of the renormalization group, we differentiate Eq.(\ref{7})
with respect to $k^2$, setting $\ka'^2 = k^2$. Then we obtain the differential equation
\be
u\frac{\partial T(u,g,\al)}{\partial u} = \ga_F(\gb^2,\bar\al)T(u,g,\al) ,
\label{11}
\en
where $u = k^2/\ka^2,~ k^2\leq0$ and $\gb^2 = \gb^2(u,g)$, $\bar\al = \bar\al(u,g,\al) = 
\al R^{-1}(u,g,\al)$ and $\ga_F(g^2,\al)\equiv u\frac{\partial T(u,g,\al)}{\partial u}|_{u=1}$,
$\ba(g^2)\equiv u\frac{\partial \gb^2(u,g,\al)}{\partial u}|_{u=1}$. It is convenient 
to introduce $\gb^2(u,g)$ as a new variable. With $
\ba(\gb^2)\equiv u\frac{\partial \gb^2}{\partial u}$, we write
\be
T(\gb^2;g^2,\al) = T\left(\frac{k^2}{\ka^2},g,\al\right) ,
\label{12}
\en
and obtain the equation
\be
\frac{\partial \ln T(\gb^2;g^2,\al)}{\partial\gb^2} = \frac{\ga_F(\gb^2,\bar\al)}{\ba(\gb^2)} .
\label{13}
\en
Here we have used
$\ab(u,g,\al) = \ab(\gb^2;g^2,\al) = \al R^{-1}(\gb^2;g^2,\al)$.
If we consider $R(\gb^2;g^2,\al)$ as given, we can write the solution in the form
\be 
T(\gb^2;g^2,\al) = \exp\int^{\gb^2}_{g^2}dx\frac{\ga_{F}(x;\ab(x;g^2,\al))}{\ba(x)},
\label{14}
\en
where use has been made of the normalization condition (\ref{6}): $T(g^2;g^2,\al) = 1$.

We are interested in the asymptotic expansion of $T(\gb^2;g^2,\al)$ 
for $\gb^2\rightarrow 0$. Because of the appearance of the gauge field structure 
function $R$ for $\al\neq0$, we have to consider separately the two cases for different 
signs of the one loop gauge-field anomalous dimension coefficient $\ga_{00}=\ga_0(\al=0)$,
and the possibility that $\gamma_{00} = 0$. For the expansion of $R(\gb^2;g^2,\al)
=R(\frac{k^2}{\ka^2},g,\al)=-k^2D(k^2,\ka^2,g,\al) $ in the limit 
$\gb^2\rightarrow0$, we have obtained in Ref.\cite{ox} the leading terms
\newpage
\be
R(\bar g^2; g^2,\alpha) &\simeq&  C_{R} (\bar
 g^2)^\xi +
 C_{R} \beta^{-1}_0 \left(\gamma_{10} - \gamma_{12}\alpha_0^2 -
 \frac{\beta_1}{\beta_0}\gamma_{00}\right) (\bar g^2)^{\xi + 1} +\cdots\cr
 &~&~~~~+ \frac{\alpha}{\alpha_0} + \frac{\alpha}{\alpha_0} \frac{\gamma_1 (\alpha_0)}{\beta_0}
 \frac{1}{1-\xi} \bar g^2 + \cdots,
\label{15}
\en
where $\alpha \geq 0$ and $\xi\equiv\ga_{00}/\ba_0 \neq 0 $ has been assumed.
The coefficients are defined by $\al_0 = -\gamma_{00}/\gamma_{01}$,  
$\beta (g^2) = \beta_0 g^4 +\beta_1 g^6 + \cdots $ , 
$\gamma (g^2,\alpha) = \gamma_0 (\alpha) g^2 + \gamma_1 (\alpha) g^4
+\cdots $, with 
$\gamma_0 (\alpha) = \gamma_{0 0} + \alpha \gamma_{01} $ ,
$\gamma_1 (\alpha) = \gamma_{10} +
\alpha \gamma_{11} + \alpha^2\gamma_{12}$ .
The special case $\gamma_{00}=0$ will be discussed later. 
For $\al=0$, the coefficient $C_R(g^2,\al)$ can be written in the form
\be
C_R(g^2,0)=(g^2)^{-\xi}\exp\int^0_{g^2}dx\left(\frac{\ga(x,0)}{\ba(x)}-
\frac{\ga_{00}}{\beta_0x}\right) ,
\label{16}	
\en
and we have $C_R(g^2,0) > 0$. For $\al>0$, we consider only parameters
$(g^2,\al)$ such that $C_R(g^2,\al)>0$. 
In this paper, we discuss non-negative values of the gauge parameter
$\al$. For these values of $\al$, the transition of exact expressions to the 
perturbative weak coupling limit is possible. Negative values of $\al$
have been considered by Nishijima in a nonperturbative framework \cite{nn}. 
Distribution aspects of the structure functions have been discussed in \cite{oz}.

\bigskip
\bigskip

${\bf \gamma_{00}/\beta_0 < 0} $ :  Let us first consider the case 
$\xi\equiv\frac{\ga_{00}}{\ba_0} < 0$, corresponding to  $\ga_{00}>0$, $\ba_0<0$.
We assume here that $\alpha \geq 0$. For QCD, 
we have $\xi = \frac{\frac{13}{2}-\frac{2}{3}
N_F}{11-\frac{2}{3}N_F}$, so that $\xi<0$ corresponds to $10\leq N_F\leq16$, 
where $|\xi|<1$ for $N_F\leq13$ and $|\xi|>1 $ for $14\leq N_F\leq16$. 
\footnote{For $N=1$ 
$SU(N_C)$ SUSY gauge theory in the Wess-Zumino representation, we have $\xi<0$ for 
$\frac{3}{2}N_C<N_F<3N_C, |\xi|<1$ for $\frac{3}{2}N_C<N_F<\frac{9}{4}N_C$ and 
$|\xi|>1$ if $\frac{9}{4}N_C<N_F<3N_C.$ Here $N_F$ is the number of flavors in 
the regular representation.}

Using Eq. (\ref{15}), we expand the integrand in Eq. (\ref{14}) as an asymptotic 
series for $\gb^2\rightarrow 0$. Keeping only essential terms. Since
$\ga_{F00}=0$ for QCD,  we find
\be
\frac{\ga_F(\gb^2,\ab)}{\ba(\gb^2)} &\simeq& \al\frac{\ga_{F01}}{\ba_0}
C_R^{-1}(\gb^2)^{-1-\xi}
- \al\frac{\ga_{F01}}{\ba_0}\frac{\al}{\al_0}C_R^{-2}(\gb^2)^{-1-2\xi}+\cdots
\nonumber \\ & & + \frac{\ga_{F10}}{\ba_0} + \cdots .
\label{17}
\en
With $\xi<0$, the number of singular terms depends upon the magnitude $|\xi|$. 
But all terms are integrable at the origin, and we obtain for 
the asymptotic expansion of $T$:
\be
T(\gb^2;g^2,\al) &\simeq& T(0;g^2,\al)\left\{ 1 -\al\frac{\ga_{F01}}
{\ga_{00}}C_R^{-1}({\gb}^2)^{-\xi} \right. \nonumber \\
& & - \al^2 \frac{\ga_{F01}\ga_{01}}{2\ga_{00}^2}\left(
1-\frac{\ga_{F01}}{\ga_{01}}\right)
C_R^{-2}({\gb}^2)^{-2\xi}  + \cdots \nonumber \\
& & \left. + \frac{\ga_{F10}}{\ba_0}{\gb}^2 +\cdots \right\},
\label{18}
\en
where terms of order $(\gb^2)^{1-\xi}, (\gb^2)^{-3\xi}\cdots$ have not 
been written out. We see that, for all $\xi<0$, we have a finite limit 
$T(0;g^2,\al)$ for $\gb^2\rightarrow0$. For the approach to $ T(0;g^2,\al)$,
which is given by Eq.(\ref{14}) with $\gb^2=0$, the one-loop 
$(\gb^2)^{-\xi}$-term is relevant for $|\xi|<1$, otherwise the two 
loop $\gb^2$-term involving $\ga_{F10}$.

So far, we have obtained the asymptotic behavior of the function $T$ for 
$\gb^2\rightarrow0$ provided $\xi<0$. With $T=-k^2B(k^2)$, we obtain then for 
$k^2\rightarrow-\infty$ along the negative real $k^2$-axis:
\be
-k^2B(k^2,\ka^2,g,\al)&\simeq& T(0;g^2,\al)\left\{1-\al\frac{\ga_{F01}}{\ga_{00}}
C_R^{-1}\left(-\ba_0\ln\frac{k^2}{\ka^2}\right)^{\xi} +\cdots\right.\nonumber\\
& & \left.+\frac{\ga_{F10}}{\ba_0}\left(-\ba_0\ln\frac{k^2}{\ka^2}\right)^{-1}+\cdots\right\} ,
\label{19}
\en
where only the essential terms have been written out. In order to obtain the 
asymptotic properties for all directions in the complex $k^2-$plane, 
we use Eqs. (\ref{9}) and (\ref{10}), and find that Eq.(\ref{19}) is also valid 
for general directions. In view of the bound (\ref{19}), we can write an un-subtracted 
dispersion representation for $B(k^2)$ :
\be
B(k^2) = \int^\infty_{-0}dk'^2\frac{\rho_B(k'^2)}{k'^2-k^2}.
\label{20}
\en
For the leading asymptotic term of the discontinuity $\rho_B$, we obtain then for 
the case $\xi<0$ under consideration,
\newpage
\be
-k^2\rho_B(k^2,\ka^2,g,\al)&\simeq&T(0;g^2,\al)\left\{-\ga_{F01}
C_R^{-1}\al\left(-\ba_0\ln\frac{k^2}{|\ka^2|}\right)^{-1+\xi} +\cdots\right.\nonumber\\
& & \left.+\ga_{F10}\left(-\ba_0\ln\frac{k^2}{|\ka^2|}\right)^{-2}+\cdots\right\} .
\label{21}
\en
Since $-k^2B(k^2)-T(0;g^2,\al)$ vanishes at infinity, 
the difference is superconvergent, and we obtain the sum rule
\be
\int^\infty_{-0}dk^2\rho_B(k^2,\ka^2,g,\al) = T(0;g^2,\al) .
\label{22}
\en
It is important to emphasize, that equal-time commutation relations have {\it not} been
used in our arguments. Sum rules similar to Eq.({\ref{22}) are usually obtained
in field theory on the basis of these  relations and with additional assumptions \cite{kl}. 
But equal time limits are very delicate in general field theories \cite{ku}.

We conclude that, for $\xi<0 ~~(10\leq N_F\leq16$ in QCD), the structure function 
$-k^2B(k^2)$ approaches a constant for all $\al\geq0$. Note that all our formulae 
are valid for $\al=0$, where the situation simplifies considerably, because there is
no dependence of the renormalization group equations upon the gauge field propagator.
In the Landau gauge $\al=0$, the asymptotic expressions (\ref{19}) and (\ref{21}) are 
independent of the parameter $\xi$. 

We complete the discussion of the case $\xi<0$ with the asymptotic expressions for 
the function $A(k^2)$ defined in Eq.(2). In all directions of the $k^2$-plane, we find 
with Eq.(\ref{8}), and in analogy to Eq. (\ref{9}),
\be
S\left(\frac{k^2}{\ka^2},g,\al\right) = T\left(\left|\frac{k^2}{\ka^2}\right|,g,\al\right)
S\left(e^{i\phi},\bar g,\bar\al\right) ,
\label{23}
\en
with $\gb$ and $\ab$ as functions of $|k^2/\ka^2|$ as given below Eq.(\ref{9}). 
For $k^2\rightarrow -\infty$, we have $\ab\rightarrow0$ for the case $\xi<0$, 
and $\gb^2\simeq(-\ba_0\ln\frac{k^2}{\ka^2})^{-1}$. Since we have assumed 
that there are no mass parameters in the action, the coefficient $S$ on the right 
hand side of Eq. (\ref{23}) should vanish for $\gb^2\rightarrow0$. The details of
this limit depends upon the specifics of possible mass generation, which we 
do not discuss here. We simply 
assume that $S(\frac{k^2}{\ka^2},g,\al)\simeq(g^2)^\lambda S_0(\al)$ for 
$g^2\rightarrow0$, where we consider $\lambda\geq1$. With this Ansatz,
we obtain from Eq. (\ref{23}), in the limit $k^2\rightarrow\infty$, the leading term :
\be
-k^2A(k^2,\ka^2,g,\al) \simeq T(0;g^2,\al)S_0(0)\left(-\ba_0 \ln
\frac{k^2}{\ka^2}\right)^{-\lambda}+\cdots ,
\label{24}
\en
and for the discontinuity we have
\be
-k^2\rho_A(k^2,\ka^2,g,\al) \simeq T(0;g^2,\al)S_0(0)(-\ba_0\lambda)
\left(-\ba_0\ln\frac{k^2}{|\ka^2|}\right)^{-1-\lambda}+\cdots .
\label{25}
\en
with $S(0)=S(\ab \rightarrow 0)$.
We again can write an un-subtracted dispersion relation for $A(k^2)$, and 
since it vanishes faster then $(k^2)^{-1}$ for $k^2\rightarrow\infty$, actually 
for any $\lambda>0$, we have the superconvergence relation
\be
\int^\infty_{-0}dk^2\rho_A(k^2,\ka^2,g,\al) = 0 .
\label{26}
\en
In the presence of mass generation, this relation  expresses the absence of 
a mass parameter in the original action \cite{ni}.

\bigskip
\bigskip

${\bf \gamma_{00}/\ba_0 > 0}$ : We now turn to the case $\xi\equiv\frac{\ga_{00}}{\ba_0}>0$,
 i.e. $\ga_{00}<0$, $ \ba_0<0$  ($0\leq N_F \leq 9$ in QCD). Here we have to 
require $\al>0$. As will be seen, the limit
$\al \rightarrow 0 $ of the asymptotic expressions does not exist in this case. 
In the range considered, we generally 
have $\xi<1$ for QCD and similar theories. In order to obtain the asymptotic terms 
for $T(\gb^2;g^2,\al)$, we have to evaluate Eq.(14), which involves $\ab=\al R^{-1}$. 
For $\xi>0$, it is more convenient to use the asymptotic expansion of $\ab$ as given
in \cite{ox}: 
\be
\ab(\gb^2;g^2,\al) &\simeq& \al_0 + C(\gb^2)^\xi + 
\frac{C^2}{\al_0}(\gb^2)^{2\xi}+\cdots \nonumber\\
& & + C\Phi(\al_0)(\gb^2)^{1+\xi}+\cdots\nonumber\\
& & + \frac{\al_0}{\xi-1}\phi_0(\al_0)\gb^2+\cdots ,
\label{27}
\en
where 
\be
& &\al_0 \equiv -\ga_{00}/\ga_{01},~~~~ C\equiv -\frac{\al_0^2}{\al}C_R(g^2,\al),\nonumber\\
& &\Phi(\al_0) = C\left(\frac{\xi+1}{\xi-1}\phi_0(\al_0)-\al_0\phi'_0(\al_0)\right),\nonumber\\
& &\phi_0(\al_0) = \frac{1}{\ba_0}\ga_1(\al_0)-\frac{\ba_1}{\ba_0^2}\ga_0(\al_0).
\label{28}
\en
Given this expansion for $\ab$, we obtain for the quark structure function:
\be
T(\gb^2;g^2,\al) &\simeq& C_T(g^2,\al)\left\{(\gb^2)^{-\frac{\ga_{F01}}{\ga_{01}}\xi}
+ \frac{C\ga_{F01}}{\ga_{00}}(\gb^2)^{-\frac{\ga_{F01}}{\ga_{01}}\xi+
\xi}+\cdots \right. \nonumber\\
&+&\left.\left( \frac{\ga_{F1}(\al_0)}{\ba_0}-  
\frac{\ba_1}{\ba_0}\ga_{F01}\al_0 
+ \frac{\ga_{F01}\al_0 \phi_0(\al_0)}{\ba_0 (\xi-1)}\right)
({\gb^2})^{1-\frac{\ga_{F01}}{\ga_{01}}\xi}+\cdots\right\}
\label{29}
\en
With $\xi<1$ and $\ga_{F01}/\ga_{01}<1$ for the theories of interest
($\ga_{F01}/\ga_{01}=8/9$ in QCD), 
only the leading term is singular for $\gb^2\rightarrow0$ and hence for 
$k^2\rightarrow\infty$. Under these circumstances, we can express the 
coefficient $C_T(g^2,\al)$ in the form
\be
C_T(g^2,\al)=(g^2)^{\frac{\ga_{F01}}{\ga_{01}}\xi}
\exp\int^0_{g^2}dx\left(\frac{\ga_F(x,\ab(x))}{\ba(x)}-
\frac{\ga_{F01}\al_0}{\ba_0x}\right) .
\label{30}
\en
With Eq.(\ref{29}), using
only the leading term, we  obtain for $B(k^2)$ in the case $\xi>0$ , $\al>0$
\be
-k^2B(k^2,\ka^2,g^2,\al) &\simeq& C_T(g^2,\al)\left(-\ba_0\ln\frac{k^2}{\ka^2}
\right)^{\frac{\ga_{F01}}{\ga_{01}}\xi}+\cdots
\label{31}
\en
in the limit $k^2 \rightarrow \infty$, and for the discontinuity
\be
\rho_B(k^2,\ka^2,g^2,\al) &\simeq& C_T(g^2,\al)\gamma_{00}
\frac{\ga_{F01}}{\ga_{01}}\left(-\ba_0\ln\frac{k^2}{|\ka^2|}\right)^
{-1+\frac{\ga_{F01}}{\ga_{01}}\xi}+\cdots .
\label{32}
\en
Since $-k^2B(k^2)$ diverges  for $\xi>0$, we have no sum rule for $\al>0$. 
But the unsubtracted dispersion representation (\ref{20}) is certainly valid also 
for $\xi>0,\al>0$. 

Let us now consider $-k^2A(k^2,\ka^2,g,\al)=S(\frac{k^2}{\ka^2},g,\al)$ 
for $\xi>0,\al>0$. We again use Eq.(23), and find for the leading term,
writing $A(k^2)$ for brevity, 
\be
-k^2A(k^2) \simeq C_T(g^2,\al)S_0(\al_0)
\left(-\ba_0\ln\frac{k^2}{|\ka^2|}\right)^{-\lambda+\frac{\ga_{F01}}{\ga_{01}}\xi}+\cdots.
\label{33}
\en
We certainly have superconvergence if $\lambda\geq1$, as we have assumed, since 
$\xi\frac{\ga_{F01}}{\ga_{01}}<1$ in the theories considered. The discontinuity
$\rho_A(k^2)$ then satisfies again the supercovergence relation (\ref{26}), and it
has the asymptotic limit
\be
-k^2\rho_A(k^2) \simeq C_T(g^2,\al)S_0(\al_0)\ba_0
\left(-\lambda +\frac{\ga_{F01}}{\ga_{01}}\xi\right) 
\left( -\ba_0\ln\frac{k^2}
{|\ka^2|}\right)^{-\lambda-1+\frac{\ga_{F01}}{\ga_{01}}\xi} + \cdots .
\label{34}
\en 

As we have pointed out, the asymptotic expression for $B(k^2)$ and $A(k^2)$ 
in the case $\xi<0$  ($10\leq N_F\leq16$ for QCD) are valid for 
the special case $\al=0$ 
(Landau gauge), wheres those for $\xi>0$  ($0\leq N_F\leq 9$) do not allow 
the limit $\al\rightarrow0$. In fact, for $\al=0$, the structure functions 
are independent of $\xi$. This parameter only enters the renormalization group 
equation (9) via $\al$, which is not renormalization group invariant parameter. 

\bigskip
\bigskip

${\bf \ga_{00}=0}$ : It remains to consider the case $\xi=0$, i.e. $\ga_{00}=0$ 
and $\ba_0<0$,  for $\al\geq 0$. For ordinary gauge theory we have $\ga_{00}=0$ for 
$N_F=\frac{13}{4}N_C$, so in QCD with $N_C=3$, it is not of direct interest. But 
it could be realized for $N_C=4$, etc. In $N=1$ SUSY SU($N_C$) theories with matter 
fields in the regular representation, we have $\ga_{00}=0$ (Wess-Zumino representation) 
for $N_F=\frac{3}{2}N_C$, and it is of interest for $N_C=$ even.

As has been pointed out, in the Landau gauge $\al=0$, we have no dependence of the 
quark propagator asymptotics upon $\xi$. Hence, for all 
values of $\xi$, the formulae (\ref{19}) for
$B(k^2)$ and (\ref{24}) for $A(k^2)$ are valid, setting $\al=0$, as are the sum
rules (\ref{22}) and (\ref{26}). Although the asymptotic limit of the gauge field 
propagator function $D(k^2)$ for $\ga_{00}=0$, $\ba_0<0$ and $\al=0$ is not needed
here, we briefly present the leading terms, because they have not been reported
elsewhere. We obtain in this case for $k^2 \rightarrow 0$ :
\be
-k^2D(k^2) &\simeq& R(0;g^2,0)\left\{1+\frac{\ga_{10}}{\ba_0}\left(-\ba_0
\ln \frac{k^2}{\kappa^2} \right)^{-1} + \cdots \right\} ,
\label{35}
\en
with $R(0;g^2,0) = \exp\int^{0}_{g^2} dx \frac{\ga(x,0)}{\ba(x)}$. We have
the sum rule
\be
\int^{\infty}_{-0} dk^2 \rho(k^2,\ka^2,g,0) = R(0;g^2,0) .
\label{36}
\en

In order to obtain the asymptotic terms for $B(k^2)$ and $A(k^2)$ 
in the case  $\xi=0$ and $\al>0$, we need the properties of the gauge field 
propagator for this case, which also has not been given before. 
For $R(\frac{k^2}{\ka^2},g,\al)=R(\gb^2;g^2,\al)=-k^2D(k^2,\ka^2,g,\al)$, 
we have the asymptotic terms
\be
R(\gb^2;g^2,\al) &\simeq& \frac{\ga_{01}\al}{\beta_0}\ln\gb^2 + 
C_{R0}(g^2,\al) + \cdots ,
\label{37}
\en
where we know that $C_{R0}(g^2,\al)=R(0;g^2,0)$. It follows that
\be
-k^2D(k^2) &\simeq& -\frac{\ga_{01}\al}{\beta_0}\ln\ln\frac{k^2}{\ka^2}+\cdots .
\label{38}
\en
The asymptotic limit of the discontinuity is given by
\be
-k^2\rho(k^2) &\simeq& -\frac{\ga_{01}\al}{\beta_0}\left(\ln\frac{k^2}
{|\ka^2|}\right)^{-1}+\cdots .
\label{39}
\en
For $\al>0$, $D(k^2)$ decreases less fast than $k^{-2}$, and hence there is no
sum rule.
Once we have the asymptotic terms of the gauge field propagator $D(k^2)$, 
we can evaluate Eq.(\ref{12}) for $T(\gb^2;g^2,\al)$. Keeping only the leading 
term, we find for $\xi=0$ and $\al>0$ in the limit $\gb^2\rightarrow0$:
\be
T(\gb^2;g^2,\al) \simeq C_0(g^2,\al)(\ln\gb^2)^{\frac{\ga_{F01}}{\ga_{01}}} + \cdots ,
\label{40}
\en
and hence for $k^2\rightarrow\infty$:
\be
-k^2B(k^2)\simeq C_0(g^2,\al)\left(\ln\ln\frac{k^2}
{\ka^2}\right)^{\frac{\ga_{F01}}{\ga_{01}}}+\cdots ,
\label{41}
\en
For the discontinuity $\rho_B$, we obtain, for real $k^2\rightarrow\infty$,
\be
-k^2\rho_B(k^2) &\simeq& -C_0(g^2,\al)\frac{\ga_{F01}}{\ga_{01}}\left(\ln\frac{k^2}
{|\ka^2|}\right)^{-1}
\left(\ln\ln\frac{k^2}{|\ka^2|}\right)^{\frac{\ga_{F01}}{\ga_{01}}-1}+\cdots .
\label{42}
\en
The coefficient $C_0$ in these relations can be expressed in the form
\be
C_0(g^2,\al) = (\ln g^2)^{-\frac{\ga_{F01}}{\ga{01}}} \exp \int^0_{g^2} dx
\left( \frac{\ga_F(x,\bar{\al} (x))}{\ba(x)} - \frac{\ga_{F01}}
{\ga_{01}}\frac{1}{x\ln x}\right) .
\label{43}
\en
Since $-k^2B(k^2)$ diverges for $k^2\rightarrow\infty$, there is no sum rule.

In order to obtain the asymptotic terms of the function $A(k^2)$ in the 
case $\xi=0$, $\al=0$, 
we can again use Eq.(\ref{23}). With our assumption for the weak coupling limit
of $A(k^2,\ka^2,g,\al)$ in Eq.(\ref{24}), we find
\be
-k^2A(k^2)\simeq C_0(g^2,\al)S_0(0)\left(-\ba_0\ln\frac{k^2}
{\ka^2}\right)^{-\lambda}
\left(\ln\ln\frac{k^2}{\ka^2}\right)^{\frac{\ga_{F01}}
{\ga_{01}}}+\cdots,~
\label{44}
\en 
and for the discontinuity,
\be
-k^2\rho_A(k^2)\simeq -C_0(g^2,\al)S_0(0)\ba_0\lambda\left
(-\ba_0\ln\frac{k^2}{|\ka^2|}\right)^{-\lambda-1}
\left(\ln\ln\frac{k^2}{|\ka^2|}\right)^{\frac{\ga_{F01}}
{\ga_{01}}}+\cdots.~
\label{45}
\en 
We see that $A(k^2)$ again vanishes at infinity faster than $k^{-2}$, 
so that the superconvergence relation (\ref{26}) remains valid for  
$A(k^2)$ with $\xi=0, \al>0$.

\bigskip

Finally we summarize the leading asymptotic terms for the structure 
functions of the quark propagator, and also include those of the 
gluon propagator, in particular for the case $\ga_{00}=0$, which has
not been given before. We use the following short-hand notation:
$\xi\equiv\frac{\ga_{00}}{\ba_0}, ~~  \al_0\equiv-\frac{\ga_{00}}{\ga_{01}} ,
~~~\ga_{F00}=0 ; ~~~
-k^2D(k^2,g^2,\ka^2.\al)=R(\frac{k^2}{\ka^2},g,\al)=
R(\gb^2;g^2,\al)=R(v), ~~~\gb=\gb(v ,g^2) $, $v=\frac{k^2}{\ka^2}$ , ~ 
with corresponding relations for $B$ with $T$ and $A$ with $S$ respectively.
We write $C_R=C_R(g^2,\al)$, and similarly for $C_T$ as well as $C_0$ ;   
$T(0)=T(0;g^2,\al)$, $R(0)=R(0;g^2,\al)$. $\rho=\rho(k^2,\ka^2,g,\al)$, 
$\rho_B$ and $\rho_A$ are the discontinuities of the corresponding structure
functions.

\ben
\xi<0: &~& \ga_{00}<0, ~\ba_0<0, ~ (13N_C< 4N_F<22 N_C), 
~~\al\geq0 \\
& &  \\
& &R(v)\simeq C_R(-\ba_0\ln v)^{-\xi} +\cdots \\
& &T(v)\simeq T(0) + \cdots \\
& &S(v)\simeq T(0)S_0(0)(-\ba_{0}\ln v)^{-\lambda} + \cdots  \\
& &\int^\infty_{-0}dk^2 \rho_B = T(0),~~~ \int^\infty_{-0}dk^2 \rho_A=0 , ~~
(\lambda > 0). 
\enn

In the case $\al=0$, the results given above are valid for all $\xi \neq 0$.
 
\ben
\xi>0: &~& \ga_{00}>0, ~\ba_0<0, ~ (0< 4N_F<13 N_C), ~ \al>0   \\
& &   \\
& & R(v)\simeq \frac{\al}{\al_0} + C_R(-\ba_0\ln v)^{-\xi} + \cdots \\
& & T(v)\simeq C_T(-\ba_{0}\ln v) ^{\frac{\ga_{F01}}{\ga_{01}}\xi} +\cdots\\
& & S(v)\simeq C_TS_0(\al_0)(-\ba_{0}\ln v)^{\frac{\ga_{F01}}{\ga_{01}}\xi-\lambda} 
 + \cdots\\
& & \int^\infty_{-0}dk^2 \rho = \frac{\al}{\al_0} , ~~~\int^\infty_{-0}dk^2 \rho_A=0 ,
 ~~(\lambda \geq 1) 
\enn

\ben
\xi=0: &~& \ga_{00}=0, ~\ba_0<0, ~~ (4N_F=13 N_C), ~~ \al>0    \\
& &     \\
& & R(v)\simeq -\al\frac{\ga_{01}}{\ba_0}\ln\ln v  +\cdots  \\
& & T(v)\simeq C_0(\ln\ln v) ^{\frac{\ga_{F01}}{\ga_{01}}} + \cdots \\
& & S(v)\simeq C_0(\ln\ln v)^{\frac{\ga_{F01}}{\ga_{01}}}(-\ba_{0}\ln v)^{-\lambda} + \cdots \cr 
& & \int^\infty_{-0}dk^2 \rho_A=0,  ~~~(\lambda > 0)
\enn

\ben
\xi=0: &~& \ga_{00}=0, ~\ba_0<0, ~~ (4N_F=13 N_C), ~~ \al=0   \\
& &     \\
& & R(v)\simeq R(0)\left(1 + \frac{\ga_{01}}{\ba_0}(-\ba_0\ln v)^{-1} + \cdots \right) \\
& & \int^\infty_0 dk^2 \rho = R(0)
\enn

With $\xi=0$, $\al=0$, the asymptotic terms for $T(v)$ and $S(v)$ are obtained by setting $\al = 0$ 
in the case $\xi < 0$ given above.  


\end{document}